# From novel mathematics to efficient algorithms. Do we have proper SD foundation to build future?

Fyodor V. Tkachov

INR RAS, Moscow 117312 Russia; University of Alberta, Edmonton, T6J 2J1 Canada

Computer implementation of sophisticated algorithms for physics applications is greatly facilitated by the new generation of component-oriented SD technologies pioneered by Wirth's Oberon and embraced by the software industry (Sun's Java and Microsoft's C#). In a stark contrast with the deeply flawed C++, the Oberon technologies (specifically, the Component Pascal and the BlackBox development environment) offer an unparalleled SD platform for building scientific applications from sophisticated data processing to demanding symbolic calculations. Specific examples are discussed.

This workshop is taking place against the backdrop of a theoretical crisis with calculations for the existing and future colliders; theorists are behind experimentalists precision-wise, and it is not clear when (and if) the gap will be significantly narrowed. The theory for LEP1 [1] was implemented within the framework of the calculational paradigm based on the use of Schoonschip and derivatives [2] for vector and spinor algebra; dilogarithms for (one loop) integrals [3]; and FORTRAN for numerical calculations; with the different pieces connected [4] by a tremendous amount of hand work. But already the one-loop calculations for LEP2 are far from being complete. What about 2 loop calculations in the Standard Model that are needed for theoretical numbers to match data precision-wise?

The purpose of this workshop as explicitly set forth by the organizers in the first bulletin was supposed to be

> “to set up the bases for a *more coherent and professional* approach of our activities both at the theoretical as well as the technical level.”

To appreciate what *more coherent* means here, consider the fact that this workshop has seen presentations from the GRACE [5], CompHEP [6], OMEGA [7], and GiNaC [8] projects — all featuring different and incompatible symbolic algebra engines, without any way to share application-specific algorithms except by rewriting source code. Clearly *coherence* between all these projects is lacking to a worrying degree.

*Professional* is, in essence, about efficiency: it is the efficiency in performing a task that differentiates professionals from amateurs. Most importantly, a professional is seen by the tools s/he uses.

It is perfectly obvious that many of the problems theorists are encountering result from a lack of a proper common SD foundation for the algorithm design work. Yet surprisingly little has been said about this at this workshop: D. Perret-Gallix mentioned the problem of standards in his introduction, and A.Kryukov discussed a technology that could help eliminate redundant effort in field theory model building [9]. And that was it.

The present talk argues that:

- physicists as a community are actually well behind the state of the art in software engineering; the circumstance results in a huge and continuing waste of resources, bound to continue into the future;
- the state of the art in SD is not the amateurish and deeply flawed C++ [10] but Oberon-2 [11] (the best supported implementation known as Component Pascal [12]);
- the Oberon[i] technologies have ushered in what may be called the modern standard SD paradigm; the paradigm encompasses the two dominant mega-projects of the software industry: Sun’s Java and Microsoft’s .NET;
- but Oberon-2/Component Pascal, while capable of peaceful coexistence with both Java and C#, remains the best foundation for complex scientific applications.

## Three causes of the calculational crisis

First, there is the problem of politics around research fund allocation, and the research community has not done enough to minimize its adverse effects on the efficiency of research. Politics means desinformation (which spans the entire range from omissions to plagiarism), and so “stifles communication, breeding distrust and inefficiency …” — in the final respect, causing a crippling misallocation of research funds[ii]. At a technical level, the resource allocation politics obstructs ***component software***[iii]: why should one bother to make public a software component for, say, fast Dirac trace calculations if there is no credit to be gotten for it?

Second, the theoretical community’s command of mathematical methods seems to be below what’s required by perturbative quantum field theory — not in regard of combinatorial complexity but in regard of the conceptual framework. It is profoundly misleading to regard Feynman diagrams as ordinary integrals; they are generalized objects and should be treated as such, with full understanding of the concepts and methods of distribution theory (I discussed this in [13]). Generalized solutions, regularization (in the sense of numerical methods), etc. are key concepts here.

---

[i] For simplicity, I will not explicitly distinguish Oberon (1986) and Oberon-2 (1991), and will always have in view the latter.

[ii] One can invent calculational methods on which a whole calculational industry would thrive, with the results receiving thousands of citations — and then be told that one “is not doing any physics” and be denied an essential support.

[iii] Component software implies that one can replace a module encapsulating, say, an algorithm for doing Dirac traces and conforming to pre-defined interface specifications, in one’s system bypassing unnecessary encumbrances like the linking step and maintenance of header files, and immediately start using the module without rewriting/recompiling/relinking one’s application program written to the same specifications. This concept is implemented in Oberon, Java and C# — but is impossible with FORTRAN, C and C++.

Examples of constructive solutions of this kind from my own experience are: the asymptotic operation [14]; the gauge-invariant perturbation theory with unstable fields [15]; algebraic (a.k.a. integration-by-parts) algorithms for multiloop calculations [16]; quasi-optimal observables [17].

But how to implement all that wonderful mathematics as working algorithms? Quite often, this requires much more experimentation and safe flexibility of data structures than what's offered by the dominant SD platforms (FORTRAN, C/C++). In other words, the dominating SD platform imposes severe penalties on the algorithmic design.

One might say that this is what division of labor is for. However, there is an objective and a subjective objection to this. Subjectively, as experience proves, whoever controls the software tends to dictate the rules of collaboration, which was seen to be disastrous in the long run. Objectively, experimenting with sophisticated algorithms may require a much more rapid cycle and broader-band feedback than what's usually possible within a team of specialists; in other words, the best results are achieved when formulae and their algorithmic implementation co-evolve in the same head. Due to these reasons I had been for a lo-o-ong time seeking simple yet powerful SD tools that would allow me to stay in control of my algorithms, intellectually and otherwise.

The third cause of the calculational crisis is a grossly inadequate SD foundation. Briefly put, physicists' SD, on the whole, is in the stone age.

- The computer revolution evolves too fast, affecting too much. This by itself creates myths. Physicists are more susceptible to such myths because they have physics to worry about, full time.

- The speed of computer revolution exacerbates the phenomenon of *effective incompetence of scientific elite.* (It is hard to stay on the cutting edge even in one's field of specialization. A step or two up in the hierarchy, and it becomes impossible to have one's own expertly opinion on the range of subjects one is supposed to supervise. Another step up, and one no longer really needs to be an expert in anything but public relations.)

- Economic principles do not work as elsewhere because it is hard to match the "product's" value against investment.

- So there is platform fragmentation: historically, physicists have been able to justify expensive hardware which carries higher profit margins for manufacturers thus weakening competition in this segment and resulting in larger array of incompatible hardware in use.

- But collaborations are essential, so physicists are essentially reduced to a crippling common SD denominator (FORTRAN, UNIX, C/C++, TeX, ROOT …), providing ground for commercial — but still inadequate — projects such as Form-2.

Consider TeX as an example of the irrational mythology that justifies the situation. 30% of the human brain is involved in the processing of visual information. Using a non-visual equation editor is like lobotomizing oneself by 1/3.[i] Yet the complexity of TeX tends to be (unconsciously) regarded as a source of pride by TeXnicians. (A similar sentiment seems to prevail in regard of C++.) I personally do not feel I have enough brain power to waste with TeX. On the technical side, TeX is a one way street: it was designed with, essentially, the sole goal in view (eliminating the typesetter by placing the burden on the author), and it is too complex to be a standard for exchange of convertible mathematical information. Even for exchange of papers postscript proves to be a preferred solution. (Most people seem to prefer to download postscript from the arXive to avoid the hassles of TeX. On the other hand, the emerging MathML standard addresses the need of mathematical information exchange with not just form but also meaning preserved [18].)

- Then there is this problem of monolithic software incapable of genuine extension (see [8] for similar arguments): try to implement a fast bit-manipulating algorithm in Maple or Schoonschip (an example is the algorithm for evaluation of Dirac traces, see below).

At this point I note that the variety of efficient algorithmic solutions needed for advanced pQFT calculations spans the entire spectrum from old-fashioned numerical calculations to dynamical Lisp-like data structures to bitwise manipulation usually considered to be a feature of system-level programming. One needs genuine procedures with various parameter-passing mechanisms, etc. In short:

> *One needs the full power of a general-purpose system programming language.*

One also needs to be able to mix software components responsible for different aspects of the calculation in an arbitrary fashion, in as easy and error-safe fashion as possible.

I won't discuss the dying FORTRAN monster. So:

### What's wrong with C/C++?

Two things are wrong with these popular languages:

(i) Programmers' **productivity** is <u>unnecessarily</u> low.

(ii) The resulting software's **quality** is <u>unnecessarily</u> low.

(We are talking here about a BIG factor, not just a few %, and about larger projects where these effects are manifest.)

There are mountains of evidence for this. The problem is, however, that physicists on the whole are ill-informed about these things. Here are some easily available studies: [19], [10]. The quotes below are from those publications.

### C

- "… the C enthusiasts seem to be largely in ignorance of the advances which have been made in language design in the last 20 years …"
- C offers no safety to the programmer: it is far, far too easy to create hard-to-find bugs (recall how Form crashes unless its author participates in the project). Remember that the main source of errors is not the initial coding (it is this initial stage that many programming gurus refer to when proclaiming that it does not really matter which language they write in), but the code's modification (e.g. to experiment with an algorithm, or to adapt code to evolving problems, etc.).

[i] At this point a member of the audience objected citing the difficulties he experienced fixing a picture on a page in "Word". Well, first, I bet the person involved did not spend on the "Word"'s documentation a fraction of the time he was forced to do with TeX. Second, there is absolutely no contradiction between the ease of visual design and the precision of manual specification: in my obsolete version of "Word", Format → Frame brings up a dialog with several options for manual specification of the picture's position (Relative To: Page, Margin, Column, Paragraph).

- Emphasis on explicit pointer arithmetic with no automatic memory management places too hard a burden on programmers and is the single most serious source of extremely hard-to-find bugs in medium- and large-scale programs with dynamic data structures.

- There are no genuine modules, which prevents a true encapsulation of code and thus preventing true component software.

The above three points also imply that production of medium- and large-scale software is greatly (and unnecessarily) impaired by the use of C.

- Poor readability — cf. the Obfuscated C Code Contest [20]. Poor readability translates into a support nightmare (or guaranteed salaries for the authors of the code).

- Then there is this ***Law of saturation of programming language's degrees of freedom***: in a large project participants will tend to use — deliberately or as a result of natural code evolution through patches etc. — all the features of the language, which inevitably leads to obscure errors when code modifications are attempted.

The law comes into a full effect: (1) with novice programmers wanting to show off their expertise (I thank T.Ohl for sharing a story); (2) whenever the project is to incorporate substantial independently-written pieces of code.

- The poor readability of C and the Law of saturation … make communication extremely difficult — thus impairing collaborations.

Also consider the following example: You are a physicist who doesn't do programming day in, day out. Suppose you wrote some code last year, then did some theory for several months to improve your formulae, and now want to modify that code. How much time would you need to regain the ability to read your C code fluently?

"… Anyone who has practiced C will know how many traps there are to fall into …" [19].

## C++

- <u>No genuine programming language design expertise</u> behind it: ".. much of the C++ literature has few references to external work or research.." [10]

- <u>A false promise of compatibility</u> with C: it is simply impossible to have a sound object/component model with explicit unchecked pointer manipulation (an independently written piece of code dropped into your project can ruin everything).

- The language is far too complex, unnecessarily complex ➔ poor portability, bad compilers.

- C++'s features such as operator overloading sharply increase dimensionality of the space in which the above Law of saturation plays out. "C is not applicable for large scale production … C++, however, has not solved C's flaws,.. but <u>painfully magnified them</u>." [10].

- C++ was an <u>unproved, untested</u> technology when adopted as a standard for large projects such as LHC.

Software industry scurries away form C++: according to the industry analysts, in 5 years Java and C# would dominate in the software industry at the expense of C/C++ [21]. Won't it happen that by mid-LHC C++ will be a dying species like FORTRAN is now?[i]

In short, adopting C++ for SD in physics now appears to be nothing short of a major disaster. "Physicists would have been better off with FORTRAN."

## What would proper SD foundation be like?

Here is a laundry list of desirable features in no particular order:

- Suitability for **standard numerical** applications; this implies efficient compiled code.
- **Symbolic algebra.** As I said already, the variety of symbolic algebra problems is huge and spans anything from dynamic Lisp-like data structures to database-like features to bit-wise combinatorial manipulation.
- **Everything in between**: there must be no hard and fast boundary between numerical and symbolic calculations. Features like FORTRAN output of symbolic algebra systems are props, not real solutions. In fact, design of advanced numeric algorithms is greatly facilitated if dynamic data structures are well supported, blurring the boundary between the two classes of applications.
- Support of **graphics** is necessary. Interactive graphics (not just dialogs) is a whole new dimension in scientific computing.
- Connection with **legacy software** (e.g. FORTRAN libraries).
- A proper SD platform must support **extensibility**: for instance, no closed SA systems can possibly provide all one needs, whence a tendency to build SA systems using general purpose languages with full access to the underlying data structures [6], [7], [8]; see also on the BEAR project below.
- Support of **collaborations** — i.e. allow independent development of plug-and-play components. This implies at least two things:
- **Modularity** (not the fake modularity of C/C++ but the true modularity allowing true information hiding as in Modula-2, Oberon, etc.).
- Full **safety** — i.e. a minimal dimensionality of the space in which the Law of saturation etc. plays out; or, equivalently, a maximal robustness of the language with respect to human errors, esp. those induced by modifications etc. (This requirement was understood and implemented by Wirth in Oberon but not by Stroustrup in C++.) A full safety implies:

— strict static type safety;
— strict control of interfaces;
— centralized memory management (this implies that no full compatibility with C, Pascal and similar languages which allow a direct manipulation of pointers is possible).

- To support collaborations, it may also prove necessary to find a mechanism for **properly crediting authors** of software components used in a project (a similar problem exists with commercial components and web services).

- **Simplicity**.

— If we want to remain physicists then the core language in — and on top of — which all the above is to be provided, must be simple enough in order to coexist in our

[i] During the workshop, anecdotal evidence was offered about a serious conflict in a large experimental collaboration in regard of the use of C++.

heads with physics, calculus …

— Physicists and programmers within the physics community must understand each other well.

— It must be possible to immediately resume programming after months of analytical work.

— In short: programming should be considered as basic a tool in our profession as calculus, and it ought to be possible to do ***programming in as casual a manner as we differentiate and integrate.***

- **Portability**. This problem becomes somewhat less pronounced with the growing popularity among physicists of Intel-compatible dual-boot Linux/Windows workstations, but it still exists.

## Some history

A brief reminder will help to appreciate where the solution I am going to advocate comes from.

Algol-60 introduced the concept of a general-purpose programming language with a strictly defined syntax. In the 70's, the techniques of structured programming were widely adopted. In the 80's, object technologies became popular (object technologies are a natural extension of the concept of user-defined records and a natural tool to support dynamical data structures). The 90's saw the pattern movement and intensive discussions of component software. Now all the rage in the industry are web services (objects/software components interacting over Internet).

Around '70 N.Wirth at ETH [22] created Pascal as a successor to Algol-60 (and subsequently received the Turing Award for this achievement in 1984). In '72-74, to facilitate ports of Pascal to various platforms, the pseudo-machine P-code was developed at ETHZ (cf. the resurgence of this idea in the Java bytecode and Microsoft Intermediate Language, the latter designed for efficient compilation; MSIL is an essential element of the .NET project, ensuring both portability and multi-language programming within .NET). In '79 Modula-2 was created as a successor to Pascal; it was aimed at full-scale general purpose system programming; it remains fully competitive with C as far as compiled code efficiency and low-level features are concerned, while far exceeding the latter in reliability of resulting systems due to strong typing and other safety features. In '83-85 Wirth built a fast and compact Modula-2 compiler, thus demonstrating advantages of both his key programming methodology of stepwise-refinement [23] and programming languages with carefully designed, compact formal definitions.

Of major importance was Wirth's Project Oberon [11]. In it, Wirth attempted, based on his experience with Pascal and Modula-2, to give a concrete, practical answer to the questions, *What are the essential elements of a general purpose procedural programming?* and, *Is it possible to distill such elements into a compact, simple and comprehensible language, yet retaining all the power of Modula-2?*

The result proved to be nothing short of a miracle: a smooth blend of the conventional structured procedural programming language features (loops, arrays …) with a rational subset of object technologies. In the process, the notion of ***component software*** received a practical implementation. To ensure that independently developed software components and objects do not break the whole, Wirth realized the absolute need for a centralized automated memory management, which implied a ban on accidental use of explicit pointer arithmetic and unchecked type casts, thus making it impossible to retain a full compatibility with either Pascal or Modula-2 (something that was completely missed in the design of C++). Surely low-level facilities are provided for exceptional situations (drivers etc.), as was the case in Modula-2.

The result, I repeat, was nothing short of a miracle:

- a simple[i], highly readable language (see code example in Appendix A); its formal definition fits on one page (see Appendix B), and Language Report is under 30 pages [12];
- a lightning fast single-pass compiler;
- no linking step (compiled modules are loaded dynamically on demand, with the necessary interfaces and versions checks) and no separate header files (these are automatically maintained by the system);
- the preceding two features result in an extremely flexible, quick development cycle, making one feel like working with an interpreted system;
- all genuine the benefits of OOP (inheritance, polymorphism) are provided without pain with only single inheritance, and with efficiency of compiled code in no way compromised;
- the language is extremely easy to learn and easy to use thanks to a strict orthogonality of its features, which leaves no room for pitfalls such as those encountered in Java (see below);
- the built-in automatic garbage collection proves to be a pure joy to live with (see also about BlackBox/Gr below) — even in situations where static memory management FORTRAN-style would be sufficient.

> There is every reason to regard Wirth's Oberon as a computer age analogue of Euclidis Elementa.[ii]

## Implementations of Oberon

Oberon is both a language and a run-time environment insofar as the prevailing OSes do not support automatic garbage collection. There are both standalone implementations of Oberon as well as Oberons running under other OSes [27]. Of the latter breed, the most stable by far and best supported is the commercial version (free for teachers and ill-funded researchers) provided by the company Oberon microsystems, Inc., [28]. For marketing reasons, the language was renamed to Component Pascal, and the development environment was called BlackBox Component Builder.

Ominc was founded in 1993 by Wirth's students with the purpose to port Oberon technologies to popular plat-

[i] Simplicity in this context does not imply that, say, more than 2-dimensional arrays, or arrays of records, are not allowed; on the contrary, each such restriction is regarded as a *complication* of the language.

[ii] At least in regard of procedural programming. However, one of the first code examples in the functional language Objective-Caml [25] that I was shown contained a FOR loop. Moreover, the compiler of O'Caml is an order of magnitude bulkier (and, presumably, slower) than that of Component Pascal. As to the formal program verification, (1) this can be done very well with structured procedural languages [26]; (2) for larger software projects formal verification is probably impossible no matter what language one uses.

forms. (NB One of the cofounders, C.Szyperski, is now at Microsoft Research.). BlackBox is currently available for both Microsoft and Apple platforms, nicely adapting to the native GUI in each case. The company's revenue is derived from consulting and custom architecture and software design with an impressive list of references. The top-level expertise of the team and the quality of BlackBox is demonstrated by the fact that Borland, the reputable maker of excellent programming tools (TurboPascal, Delphi), commissioned Ominc to write a JIT compiler for their Java VM.

The language **Component Pascal** is an Oberon-2 fine-tuned to provide an improved support for large systems and compatibility with with Java at the level of base types.

**BlackBox Component Builder** is an industrial-strength RAD IDE, featuring a unique combination of properties not found in any other similar tool on the market:

- it is fast (runs well on a i486) and compact (distribution comes in a 6 MB file);
- its stability and quality of compiler are legendary;
- the software development under BlackBox is definitely easier than in Delphi (confirmed by some of my initially skeptical students);
- the resulting compiled code is perfectly clean, noticeably better than code produced by Gnu compilers (reported by my inquisitive students);
- the construction of dialogs is as easy as with Visual Basic and Delphi;
- (interactive, real-time) graphics: nothing short of *amazing* in regard of both programming ease and power;
- full access to the native OS interfaces (also to MS Office under MS Windows);
- a direct access to hardware (registers, RAM, etc.);
- full access to legacy code (the legacy code has to be recompiled into dll's);
- it is inexpensive and even free for teachers and ill-funded researchers.

The only weak point in regard of the wish list given above is platform support: here I can only quote the BlackBox documentation where it discusses possible values of the variable *Dialog.platform* ("… indicates that BlackBox is running on …"):

windows32s, macOS "not supported anymore"
windows 95, NT4, 2000, 98
macOSX, linux, tru64 "currently not supported"

## The new standard SD paradigm and component-oriented programming

The SD paradigm ushered in by the Project Oberon [11] can be described as one based on the use of a safe, strictly typed, structured, modular programming language which incorporates a rational subset of object technologies (without multiple inheritance), with the run-time system supporting dynamic loader with version control and strict interface control as well as automatic memory management (the garbage collection well-known e.g. in Lisp and the functional programming technologies).

The Oberon paradigm has by now become the "standard SD paradigm" — thanks to the adoption of its key concepts by the Sun's Java and Microsoft's C#/.NET project:

Sun's **Java** introduced in 1995 is clearly seen (and known) to have been influenced by Oberon, the most obvious difference being the adherence to a C-style syntax. Java's intermediate bytecode also makes one recall Wirth's P-code project which ensured portability of Pascal to various hardware platforms.

Unfortunately, Java has never been designed for run-time efficiency or numerical applications. In the interpreted version, the best Java JIT compilers produce code whose efficiency does not exceed 60% of C++, and the programs written in C++ can be significantly slower than C.[i] Even if directly compiled into native machine code, Java does not allow object allocation on stack, and Java's parameter passing mechanism may cause inefficiencies (in a simple test, a model of the Java parameter passing was slower by a factor of 2 than a semantically equivalent Component Pascal program with three OUT parameters). The floating point support in Java may also be less than adequate [29]. But probably the biggest concern with Java is its unnecessary complexity resulting from un-expertly language design:

"… with Java, one is always having to revise one's "knowledge" of it as more is learnt .. incredible but hidden complexity — such as the obscure rules for inheritance … Java … looks simple yet is complicated enough to conceal obvious deficiencies." [30]

The programming language **C#** (2000) is a central part of the mega-project Microsoft.NET; the language is similar to Java in many respects including C-style syntax, otherwise the influence of Wirth's school of thought is even more pronounced:

- Microsoft Research, responsible for the design of .NET, hired C.Szyperski, a cofounder of Oberon microsystems and a leading world expert in component software, as well as the creator of Turbo Pascal A.Hejlsberg.
- Oberon-2 and Component Pascal were both among the 12 languages presented in the 2000 announcement of .NET.
- The intermediate code (MSIL) is designed for efficient compilation and thus directly compares with the Pascal P-code. There is evidence that code compiled from Component Pascal via MSIL compares with C compiled natively [31].

Unlike Java, C# has been standardized by ECMA. Also deserves a mention the Mono Project by Ximian [32]; the aim of the project is to bring the core of .NET to Linux.

However, with C# there are also efficiency concerns, although to a lesser degree than with Java: all objects in C# are allocated on heap which may in certain circumstances cause unnecessary overhead. The complexity of C# may also be an issue: although the language is very much simpler than C++, its designers failed to refrain from language design experiments ([un]boxing, etc.) [33] as well as stuffing language with features which ought to be relegated to libraries.

Of the triad, Oberon/Component Pascal obviously stands out for its simplicity, uncompromised suitability for nu-merical applications (in which regard it is as good as FORTRAN or C) as well as a clean and expertly language design, making it much easier to have a complete intellectual control over — which is a prerequisite for high-quality software development, especially by non-professional developers.

[i] C++ gurus argue that this is solely due to a poor programming. However, the evidence at my disposal (reported by my collaborators on the MILX project [40], now at Qwest Communications) concerns fairly mundane tasks. I don't see, for instance, how a straightforward program could be ported from C, say, to Component Pascal under BlackBox and become appreciably slower (if at all). Of course these observations do not constitute an accurate statistics.

## Examples

In the examples below I try to demonstrate the advantages of the Oberon paradigm in general, and the BlackBox IDE in particular, from many different angles.[i] A more technical feature-by-feature comparison is found in [34]; some of the propositions of [34] are somewhat subjective, some are too narrow (like the notion that object programming features fall outside the scope of scientific programming, which I completely disagree with), and some are obsolete (e.g. experience has confirmed that negative effects of operator overloading in large projects outweigh its syntactic convenience); however, the core argumentation and conclusions of [34] are in agreement with mine.

As a warm-up, recall that the CompHEP project [6] was originally created in the late 80's with the Turbo Pascal. Creating such a system from scratch was no mean feat, and CompHEP remains unique in some respects to this day. However, to enable its use by physicists on more "powerful" platforms, it was ported to, and rewritten in the spirit of C. Since then, its core has not seen serious enhancements despite its known and undesirable limitations.

The next example is a widely quoted testimony by a Microsoft manager [35]: "By writing in Java, Microsoft programmers are between 30% and 200% more productive than when they write in C++." This must concern the tasks of business-oriented programming. I add from experience that in the design and implementation of sophisticated mathematical algorithms, the automatic garbage collection in Oberon/Java/C# (as opposed to C++) seems to offer even greater benefits:

### Optimal Jet Finder

My next example demonstrates the advantages offered by Component Pascal and the BlackBox in regard of development of numerical algorithms of the less standard nature. This is an implementation of the so-called optimal jet definition [36], [37]. What's interesting is that a precursor of the definition was discussed 20 years ago [38] and as recently as [39] it was claimed that computer implementation of such mathematical definitions is computationally prohibitive; indeed, here one has to find a global minimum in a O(1000)-dimensional domain.

Yet the ease of experimentation with algorithms (resulting from simplicity and transparency of SD with the BlackBox) allowed to build the first version in Component Pascal in just ***4 weeks***. I emphasize that it was not a priori clear what kind of algorithm would do the trick, and in all the experimentation that had to be done, the safety features of the language helped enormously. Next, a port to FORTRAN 77 with subsequent attempts of fine-tuning (together with a modification of the code required by a modification of the mathematical definition) took ***months***. Only reverting to Component Pascal allowed to fix the problem there — in ***days***. Production of the final FORTRAN code again required an inordinate amount of time — of the order of a ***month***. The times include all the usual real-life distractions which put the longer projects at a disadvantage due to time losses for restoring context after interrupts, etc.

The first remark is that the ease of development is absolutely incomparable between old-style FORTRAN and the new-style Component Pascal. *There is just no comparison.*

The second remark is that such an algorithm *could* be a software *component* — a piece of code with an agreed-upon interface that could be simply dropped by other people into their systems and used — without all the headaches of relinking, variables' names conflicts, maintenance of header files, etc.

The third remark is that, once one gets the hang of how languages of the Oberon paradigm are to be used and learns the corresponding idiomatics, one realizes that there is a vast and usually disregarded class of numerical algorithms which make a casual use of dynamical data structures (OJF barely escaped ending up as an algorithm of that kind). Such algorithms bridge the chasm between the usual static numerical number crunching and symbolic manipulation, and offer vast new opportunities not yet fully appreciated (e.g. for sophisticated adaptive integration algorithms [40]). The development of such algorithms can be very difficult without automatic memory management as I learned from the MILX project [40].

### Dirac traces in D dimensions

This example concerns the standard problem of evaluation of Dirac traces in D dimensions — a problem that occurs in a vast number of theoretical high energy physics calculations. The example demonstrates the power of bit-manipulation facilities in Component Pascal as well as the power of flexible approach to algorithm construction offered by the Oberon paradigm. The algorithm I am going to discuss was announced on my web site in 1998 [41].

It is a funny algorithm: it involves no fancy mathematics, essentially, only optimizations of various kinds — formulaic and software. Start with simplest formula, quoted pro forma in any quantum field theory textbook:

$$\text{trace} = \sum \text{product of pairings}.$$

This is usually considered as completely unfit for algorithmic implementation due to a huge amount of calculations involved with the sum. However, take this formula seriously, optimize the combinatorics involved in a straightforward fashion (order of summations, etc.), implement it using the systematic bit-manipulation facilities of Component Pascal, optimize algorithms (I am talking here about fairly trivial optimizations such as replacing often occurring procedure calls with special parameters with in-line code, etc.). The result of this exercise turned out to be surprising indeed. Consider the trace of a string of $\gamma$-matrices of the form a1b2c3d4e5A1B2C3D4E5 where letters and numbers represent vectors and summation indices, respectively. Here is a part of the resulting output produced by Form-2 (Personal Version Tkachov):

```
Bytes used = 75566966
...
Time = 1249.89 sec
Generated terms = 5490390
```

[i] The examples have *not* been selectively compiled as a C++ fan in the audience claimed to be the case.

By contrast, the new algorithm took, on the same machine, only 117.2 sec to do the same job (i.e. 10 times faster) in only 20K (i.e. in 3000 times less memory): the new algorithm produces the result term by term within statically allocated memory. The use of a 32-bit version of Form-2 speeds the calculation up only by a factor of 2.

Admittedly, the considered trace is not the simplest one by far, but, say, in realistic 4-loop calculations with two $\gamma_5$-matrices treated as antisymmetrized products of 4 $\gamma$-matrices each, things get much worse [41].

Again, such an algorithm *could* be a *component* — and it is easy to imagine, say, a componentized Schoonschip into which the code implementing the algorithm, written to adhere to a pre-defined interface specification, could be dropped to replace a less efficient version — without relinking and other chores.

## Basic Extensible Algebra Resource

This is a toolbox to support my experimentation with various algorithms related to Feynman diagrams, primarily including — but not limited to — the algebraic algorithms for loop calculations [16]. This is in fact a compact component framework for symbolic manipulation of the kind I outlined in a wish list in [42] as a post mortem for the development effort that had lead to the famous Mincer [43]. The general approach in regard of handling very large expressions follows that of M.Veltman's legendary Schoonschip but allows rather more flexibility.

Note that projects to create custom general-purpose symbolic algebra systems continue to emerge (e.g. [7], [8]). These are manifestations of the fundamental fact that *the variety of symbolic manipulation problems is much larger than any closed proprietary system can ever provide for;* that *custom-designed data structures are key to huge gains in efficiency;* and that *a full access to the power and flexibility of a compiled general purpose programming language is a prerequisite for building efficient solutions of really complex problems.*[i] It is convenient to call them *open-guts* computer algebra systems.

BEAR differs from a typical open-guts system in one significant respect. An emphasis on solutions with custom representations of expressions makes superfluous design of a full array of symbolic entities like scalars, vectors, etc. BEAR only abstracts and implements algorithmic support for special complicated tasks such as sorts. (Of course, whenever a more or less universally useful definition emerges, it can be easily incorporated into the framework by way of the standard Oberon/BlackBox extension mechanisms.)

BEAR is being designed to support specific solutions to the specific problems my group focuses on, without universal ambitions. Its general structure (although not the implementation) is pretty obvious:

(i) Arithmetic subsystem designed for maximum efficiency in arbitrary-precision integer number crunching.

(ii) Sort subsystem offers algebraic modifications of some standard sort algorithms (different problems require different algorithms for maximal efficiency), including sorts of arbitrarily large expressions stored on disk. The modular, abstracted nature of the system lends itself to a verification [26], whereas safety features of Component Pascal ensure a superb reliability of the resulting algorithms.

(iii) Combinatorics module provides (supposedly efficient) support for multinomial coefficients, permutations, etc.

(iv) Problem-area specific modules (Dirac traces etc.).

On top of that, the user is supposed to provide problem-class specific modules (e.g. manipulation of *D*-dimensional vectors in the original 1981 integration-by-parts algorithm [16]) and problem-specific code for concrete applications. In fact, it is hard to draw a line between BEAR and the specific solutions it supports — similarly to how there is no hard and fast line between the core BlackBox system and the various supporting subsystems.

Some architectural principles in the design of BEAR are as follows:

- The framework must be recursively architectured around very few simple but universal interface and design patterns. Finding patterns that fit the bill is critical because otherwise the development of problem-specific algorithms is bogged down by the amount of coding involved in custom design of data structures.
- The use of the standard OOP techniques (separation of interface from implementation, etc.) allows one to replace, say, sort routines without affecting the problem-specific code, thus allowing to experiment with different algorithms.
- The emphasis on a maximal localization of algorithms and data structures to increase efficiency of handling very large amounts of data on hardware with a hierarchical multi-level memory architecture.
- Allowing a maximal use of compiled code in user algorithms; interpretation of the pervasive arithmetic with integer counters and indices and the corresponding logical expressions must not weigh down on the overall efficiency. This was the idea behind Schoonschip's provision for a compiled FORTRAN subroutine, here carried to the extreme. (When coupled with appropriate mathematical solutions [44], it proved to be a crucial element for making the original Mincer work; also, this is another example of the welcome blurring of boundaries between conventional static-memory number crunching and highly dynamic algorithms of symbolic algebra.)

BEAR is currently being played with in the author's group. There are no specific plans for its distribution because this would require a support and documentation effort which I have little incentive to undertake (especially in view of my past experiences with Mincer; for the same reasons, the project planning does not provide for availability for hire of knowledgeable students for another 5-8 years).

The accumulated experience with BEAR indicates that the raw efficiency gains of resulting problem-specific algorithms compared with similar algorithms running on Form-2 can be orders of magnitude. The coding of algorithms is, of course, a problem for less experienced users — but then I remember my first encounter with Schoonschip very well: efficient large-scale symbolic algebra manipulation has never been easy, whereas the programming techniques needed with BEAR do not go beyond a fairly standard toolkit of general purpose programming, with a rather less than overwhelming dose of objects and patterns. Once the basic framework is in place, the coding of, say, the widely used 1981-style integration-by-parts algorithms is much easier — and I'd say very much more enjoyable — than was the case in [43]. More important, however, is the new range of options in regard of what kind of algorithms may be efficiently implemented. Not to mention a full control over the resulting software and its applications.

[i] An example which demonstrated these points was reported in [24].

### A.Vladimirov's usability experiment

An interesting experiment was conducted a while ago by Aleksey Vladimirov[i], who possesses a good enough understanding of computer programming — yet cannot afford to be a full-time expert in, say, C++. He needed to verify some algebraic identities via "multiplication of orthogonal projectors in the Hecke algebra with 5 generators". Never mind what this means; enough to say that conventional computer algebra systems choked with intermediate expressions — a classical example of the intermediate combinatorial blowup. So Aleksey figured out an algorithm which involved a simple representation of individual terms and a tree-like data structure to do the sort — something a general-purpose language allows easily, but not the limited languages of the conventional computer algebra systems. To cater for future needs, Aleksey decided to use this problem as a benchmark to see how well different SD environments would behave out of the box (remember, we don't want to become full-time experts in any particular compiler). Here are his results:

| Language/compiler | CPU time | memory used |
|---|---|---|
| **Oberon** | | |
| *Component Pascal/BBox* | *25 sec* | *6 MB* |
| V4 | failed | heap size low |
| XDS | failed | whole overflow |
| **Pascal** | | |
| FPK (32 bit) | 20 sec | < 50 MB |
| Delphi 4 | 20 sec | > 50 MB |
| GNU (UNIX) | ~ 1 min | ~ 30 MB |
| **C++** | | |
| GNU g++ (UNIX) | 75 sec | ~ 20 MB |
| Borland 5.02 | failed | heap size low |
| MS Visual 4.2 | failed | heap size low |

All calculations were done on equivalent hardware. The timing was done by hand, so the error margin is on the order of seconds. Similarly crude are memory estimates (except for Component Pascal where precise information on memory useage is available).

The table mostly speaks for itself.

BlackBox's overall ease of use was specifically noted.

### BlackBox/Gr [46]

The last example is the BlackBox/Gr toolbox written by W.Skulski from University of Rochester. This is a sophisticated interactive experimental data acquisition and monitoring toolbox with real-time graphical histogramming, etc. For the purposes of this talk the following quote from its documentation should suffice (emphasis by F.T.):

"*… Someone will ask the following question "why did you choose the BlackBox compiler to develop the Toolbox, while there is a well-known compiler <name>, which everybody knows how to use?"*

*… my answer is as follows: I spent hundreds of hours developing Gr, and BlackBox has not crashed on me even once.*

*Gr is not entirely trivial, and potential for programming errors is huge. And indeed, I have made many programming mistakes. I dereferenced NIL pointers and I jumped out of array bounds. I unloaded a running code from memory, while the corresponding display was still on screen. I abused the environment in many different ways.* ***It never crashed.*** *I never had to worry about leaking memory.* ***I never saw the words "segmentation violation" or "core dump".*** *If you can say the same about your compiler <name>, then please tell me its name …*

*… The reader probably does not fully appreciate the great simplification that this approach is bringing. Full appreciation comes only after the BlackBox environment has been used for some time …*"

To this testimony I add that after a short time spent developing in the never-crashing BlackBox with its lightning fast compiler and dynamically (re)loaded modules, ROOT [47] — a buggy piece of software which the smart, Windoze-bashing physicists developed, have widely deployed and now rely upon — looks like a cruel joke.

### Summary

Software industry is scurrying away from the pathetically inadequate C++ (funny that industry support must have been cited when C++ was chosen to be a standard for LHC software) and converging towards the new standard paradigm of component-oriented programming pioneered by Wirth's Oberon [11] — a type safe, structured, modular, efficiently compiled, general-purpose programming language which incorporates a rational subset of OOP features without duplication of concepts, and supports architecture design of large systems. The language (its latest iteration being known as Component Pascal [12]) is small, transparent, and highly readable (allowing a complete intellectual control over one's programs), and — through a meticulous design — highly robust in regard of programmer's errors.

A continuing use of C++ is a continuing waste of valuable resources now, and an even greater waste in future due to support costs. *Is there a mechanism* ***to stop that?*** For my part, I decided not to produce a C++ version of the Optimal Jet Finder [37].

Note that the GNU Oberon compiler is freely available. It is slower than the BlackBox compiler, and does not allow one to enjoy the benefits of an excellent IDE that the BlackBox is, but would do as a starting point due to portability. If a C++ programmer cannot master Oberon/Component Pascal in a week, they should be fired. A week-long transition to Oberon/Component Pascal would free considerable resources that could be much better spent than supporting throngs of programmers currently involved in the agony of debugging C++ codes.

***Fortunately, this cannot happen*** — cerialinly not soon. So my group, along with a few savvy experts like Aleksey Vladimirov and Wojtek Skulski, will be having fun for a while:

*Seeing one's algorithmic ideas smoothly materialize in an efficient code without hitting the multitude of unnecessary and avoidable impediments, is tremendously satisfying.*

[i] Incidentally, Aleksey's trick of IR rearrangement [45] sparked what can be called the Russian multi-loop revolution, its two other key mathematical technologies being the algebraic algorithms [16] and the asymptotic operation [14].

*Acknowledgments*. I thank D.Perret-Gallix and the Japanese hosts for their hospitality during the workshop; the Oberon microsystems crew and Niklaus Wirth for several discussions; M.Veltman for a discussion of scientific programming (in 1998; after 1999 the views expressed might have been somewhat different); A.Vladimirov for providing results of his usability experiment; G.Jikia and V.Borodulin for running the Dirac trace test on a 32-bit version of Form-2; T.Ohl for insightful discussions of functional programming; A.Czarnecki for the hospitality at the University of Alberta where this text was written up. The support came in parts from Maison Franco-Japonais; Russian Foundation for Basic Research grant 99-02-18365; NATO grant PST.CLG.977751; the Natural Sciences and Engineering Research Council of Canada.

## Appendix A. Example of Component Pascal code [37]

```
MODULE  OjfinderKinematics; (** verification version  **)
  IMPORT  Log := StdLog,  Math;
  CONST eps_round = 1.0E-10; eps_norm = 1.0E-10;
    eps_Et = 1.0E-10;
  TYPE
    Vector* = RECORD  E-, x-, y-, z-: REAL  END;
    Kinematics* = POINTER TO ABSTRACT RECORD END;
    Particle* = POINTER TO ABSTRACT RECORD
      p-: Vector; (* 4-momentum *)
      theta-, phi-: REAL
    END;
    Jet* = POINTER TO ABSTRACT RECORD
      qphys-, qtilde-: Vector (*4-momentum [normalized]*)
    END;
  …
  VAR  pi180-: REAL;  cylinder-, sphere-: Kinematics;
  PROCEDURE PosProduct* ( v, w: Vector ): REAL;
    VAR  res: REAL;
  BEGIN
    res := v.E * w.E - v.x * w.x - v.y * w.y - v.z * w.z;
    ASSERT( res >= - eps_round, 60 );
    RETURN  MAX( 0, res );
  END PosProduct;
  …
END OjfinderKinematics.
```

*Comments*. An asterisk after an identifier in a declaration signifies full export (public fields); a minus in that position signifies a limited (read-only) export. By default, all other variables or other objects defined within a module are private to that module. The ASSERT statement causes the program to abort if the logical expression in the argument evaluates to FALSE. REALs are always 8 bytes (double precision). The module is loaded at the first invocation of any procedure defined in it and stays loaded in memory; the module's variables retain their values until the module is unloaded from memory, manually or using meta-programming features of BlackBox. For further details see the Language Report [12].

## Appendix B. Syntax of Component Pascal [12]

Presented below is the *entire* list of syntax rules of Component Pascal in the extended Backus-Naur notation.

```
Module       = MODULE ident ";" [ImportList] DeclSeq
               [BEGIN StatementSeq]
               [CLOSE StatementSeq] END ident ".".
ImportList   = IMPORT [ident ":="] ident
               {"," [ident ":="] ident} ";".
DeclSeq      = { CONST {ConstDecl ";" }
               | TYPE {TypeDecl ";"} |
               VAR {VarDecl ";"}}
               {ProcDecl ";" | ForwardDecl ";"}.
ConstDecl    = IdentDef "=" ConstExpr.
TypeDecl     = IdentDef "=" Type.
VarDecl      = IdentList ":" Type.
ProcDecl     = PROCEDURE [Receiver] IdentDef
               [FormalPars] ["," NEW] [","
               (ABSTRACT | EMPTY | EXTENSIBLE)]
               [";" DeclSeq [BEGIN StatementSeq]
               END ident].
ForwardDecl  = PROCEDURE "^" [Receiver]
               IdentDef [FormalPars].
FormalPars   = "(" [FPSection {";" FPSection}] ")"
               [":" Type].
FPSection    = [VAR | IN | OUT] ident {"," ident} ":" Type.
Receiver     = "(" [VAR | IN] ident ":" ident ")".
Type         = Qualident
               | ARRAY [ConstExpr {"," ConstExpr}]
                  OF Type
               | [ABSTRACT | EXTENSIBLE | LIMITED]
               RECORD ["("Qualident")"] FieldList
               {";" FieldList} END
               | POINTER TO Type
               | PROCEDURE [FormalPars].
FieldList    = [IdentList ":" Type].
StatementSeq = Statement {";" Statement}.
Statement    = [ Designator ":=" Expr
               | Designator ["(" [ExprList] ")"]
               | IF Expr THEN StatementSeq
                  {ELSIF Expr THEN StatementSeq}
                  [ELSE StatementSeq] END
               | CASE Expr OF Case {"|" Case}
                  [ELSE StatementSeq] END
               | WHILE Expr DO StatementSeq END
               | REPEAT StatementSeq UNTIL Expr
               | FOR ident ":=" Expr TO Expr
                 [BY ConstExpr]  DO StatementSeq END
               | LOOP StatementSeq END
               | WITH [ Guard DO StatementSeq ]
                  {"|" [ Guard DO StatementSeq ] }
                  [ELSE StatementSeq] END
               | EXIT | RETURN [Expr] ].
Case         = [CaseLabels {"," CaseLabels} ":"
               StatementSeq].
CaseLabels   = ConstExpr [".." ConstExpr].
Guard        = Qualident ":" Qualident.
ConstExpr    = Expr.
Expr         = SimpleExpr [Relation SimpleExpr].
SimpleExpr   = ["+" | "-"] Term {AddOp Term}.
Term         = Factor {MulOp Factor}.
Factor       = Designator | number | character | string \
               | NIL | Set | "(" Expr ")" | " ~ " Factor.
Set          = "{" [Element {"," Element}] "}".
Element      = Expr [".." Expr].
Relation     = "=" | "#" | "<" | "<=" | ">" | ">=" | IN | IS.
AddOp        = "+" | "-" | OR.
MulOp        = " * " | "/" | DIV | MOD | "&".
Designator   = Qualident {"." ident | "[" ExprList "]" | " ^ "
               | "(" Qualident ")" |"(" [ExprList] ")"}[ "$" ].
ExprList     = Expr {"," Expr}.
IdentList    = IdentDef {"," IdentDef}.
Qualident    = [ident "."] ident.
IdentDef     = ident [" * " | "-"].
```